\documentclass[12pt]{article}
\usepackage{amssymb} 
\usepackage{epsfig}
\newcommand{\be}{\begin{equation}}
\newcommand{\ee}{\end{equation}}
\newcommand{\bea}{\begin{eqnarray}}
\newcommand{\eea}{\end{eqnarray}}

\begin{document} 

\begin{center}
{\bf ON THE STATUS OF  NEUTRINO MIXING AND OSCILLATIONS}
\end{center}

\begin{center}
S. M. Bilenky 
\footnote {Report at the XVI Rencontres de Physique de La Vallee  d'Aoste,
La Thuile, Aosta Valley (Italy) March 3-9, 2002 }\\
\vspace{0.3cm} {\em IFAE, Facultat de Ciencies, Universidad Autonoma
de Barcelona, 08193, Bellaterra, Barcelona, Spain \\}

\vspace{0.3cm} {\em  Joint Institute
for Nuclear Research, Dubna, R-141980, Russia\\}
\end{center}

\begin{abstract}
Evidences in favor of neutrino oscillations, obtained in the solar and atmospheric neutrino experiments, are  discussed. Neutrino oscillations in the solar and atmospheric ranges of the neutrino mass-squared differences are considered
in the framework of the minimal scheme with the mixing of three massive neutrinos.
\end{abstract}

There exist at present convincing evidences of neutrino oscillations
obtained in the atmospheric \cite{AS-K,Soudan,MACRO} and in the solar 
\cite{Cl,Kam,GALLEX,GNO,SAGE,S-K,SNO,SNONC,SNOCC}
neutrino experiments.\footnote{The LSND \cite{LSND} indication in favor of neutrino oscillations  requires 
confirmation. The MiniBooNE experiment \cite{MiniB} at the Fermilab, which 
aim to check
the LSND claim, is going on at present. The first result is planned to be available  
in 2004.}
We will present here the theory of the neutrino oscillations and
existing evidences in favor of oscillations. We will
consider neutrino oscillations in the framework of the minimal scheme of the 
three-neutrino mixing and we will demonstrate that in the leading approximation neutrino oscillations in the atmospheric and solar ranges
of the neutrino mass-squared differences are described by
the two-neutrino type formulas.

\section {Basics of neutrino mixing and oscillations}

\begin{enumerate}

\item
\begin{center}
{\bf Interaction of neutrinos with matter; flavor neutrinos}
\end{center}
\vskip0.5cm
\begin{enumerate}
\item
Investigation of neutrino oscillations is based on the assumption,
confirmed by all existing data, that the
interaction of neutrinos with matter is given by 
the Standard Model charged current (CC) and neutral current (NC) interactions. 
The leptonic CC and NC of the SM are given by

\begin{equation}
j^{\mathrm{CC}}_{\alpha} = \sum_{l=e,\mu,\tau} \bar \nu_{lL} \gamma_{\alpha}l_{L};\,~~
j^{\mathrm{NC}}_{\alpha} =\sum_{ll=e,\mu,\tau} \bar \nu_{lL}\gamma_{\alpha}\nu_{lL}\,.
\label{001}
\end{equation}

\item
The charged current interaction determines the {\it notion of flavor neutrinos and antineutrinos}. For example, in the decay
 $\pi^{+}\to \mu^{+} + \nu_{\mu}$ 
together with $\mu^{+}$
the left-handed flavor muon neutrino $\nu_{\mu}$
is produced, flavor electron antineutrino $\bar \nu_{e}$ produces
$e^{+}$ in the process $\bar \nu_{e}+ p \to e^{+}+n $ 
etc. It is important to stress that in the case of neutrino mixing the states of flavor neutrinos are not states with definite masses.
\item
Three flavor left-handed neutrinos $\nu_{e},\nu_{\mu},\nu_{\tau}$ and
three flavor right-handed antineutrinos $\bar \nu_{e},\bar \nu_{\mu},\bar \nu_{\tau}$ exist in nature. 
From the results of the LEP experiments on the measurement of the width of the  decay $Z \to \nu + \bar\nu $ for the number of flavor neutrinos 
the value

\begin{equation}
n_{\nu_{f}} = 3.00 \pm 0.06 \,.
\label{002}
\end{equation}
was obtained\cite{PDG}.
From the global fit of the LEP data for $n_{\nu_{f}}$ it was found

\begin{equation}
n_{\nu_{f}} = 2.984 \pm 0.008 \,.
\label{003}
\end{equation}
\end{enumerate}

\item

\begin{center}
{\bf Neutrino mixing }
\end{center}
\vskip0.5cm

According to the neutrino mixing hypothesis (see \cite{BilP})  the flavor neutrino field
$\nu_{lL}$ is a unitary combination of the left-handed components of the
fields of neutrinos $\nu_{i}$ with definite masses $m_{i}$
\be
\nu_{lL} = \sum_{i=1}^{3} U_{li} \nu_{iL}\,.
\label{004}
\ee

Here 
$ U $ is a 3$\times$ 3 PMNS \cite{BP,BPon,MNS}
unitary mixing matrix.

{\em Neutrino mixing is different from quark mixing.}

In fact, in the quark case 

\begin{enumerate}

\item
There is the hierarchy of couplings
between families
$$\sin \,\theta_{12}=\lambda \simeq 0.22;\,~\sin \,\theta_{23}\simeq 
\lambda^{2};\,~\sin \,\theta_{13}\simeq \lambda^{3}\,,$$
where $\theta_{12}$,  $\theta_{23}$
and  $\theta_{13}$
are CKM mixing angles in the standard parametrization of the CKM matrix
($\sin \,\theta_{12}$ characterizes the coupling between the first and the second families etc).

\item

Masses of the quarks satisfy the hierarchy
$$ m_{d}\ll m_{s} \ll m_{b}\,.$$
\end{enumerate}

In the neutrino case
\begin{enumerate}

\item
$$\sin^{2} \,\theta_{12}\simeq 0.3;\,~
\sin^{2} \,\theta_{23}\simeq 0.5;\,~
|U_{e3}|^{2}= \sin^{2} \,\theta_{13}\ll 1\,.$$
\item
Neutrino mass-squared differences satisfy the hierarchy
\footnote{Neutrino masses are numerated in such a way that $ m_{1}\leq m_{2}
 \leq
 m_{3}$. We assume that $\Delta m^{2}_{21}\simeq \Delta  m^{2}_{\rm{sol}}$ and
$\Delta m^{2}_{31}\simeq \Delta m^{2}_{\rm{atm}}$, where
$\Delta m^{2}_{\rm{sol}}$ and $\Delta m^{2}_{\rm{atm}}$ are the values of
neutrino mass-squared differences that were obtained from two-neutrino analyses
of the solar and the atmospheric neutrino data, respectively;
$\Delta m^{2}_{i1}=  m^{2}_{i}- m^{2}_{1}$.
 }
$$\Delta m^{2}_{21} \ll \Delta m^{2}_{31}\,.$$

\end{enumerate}
The minimal neutrino mass $m_{1}$ and the spectrum of neutrino masses are unknown.
From the data of the $^{3} H$ experiments \cite{Mainz,Troitsk}
an upper bound $m_{1}\leq (2.2 -2.5)\,~ \rm{eV}$
was obtained.

{\em Neutrino mixing can be fundamentally  different from the quark mixing.}

\begin{enumerate}
\item
Quarks are charged Dirac particles. 
For neutrinos there are two possibilities.
Massive neutrinos  $\nu_{i}$ can be {\em Dirac or Majorana particles.}
If massive neutrinos are Dirac particles, in this case the total lepton number $ L$ is conserved and the lepton numbers of
$\nu_{i}$ and $\bar \nu_{i}$ are equal, correspondingly, to 1 and -1.
If there are no conserved lepton numbers, massive neutrinos 
$\nu_{i}$ are purely neutral Majorana particles (
$\nu_{i}\equiv \bar \nu_{i} $.)
\item
The number of the massive neutrinos can be larger than the number of flavor neutrinos (three). For 
 the mixing we have in this case
\begin{equation}
\nu_{lL} = \sum_{i=1}^{3+n_{s}} U_{li} \nu_{iL}\,~~
\nu_{sL} = \sum_{i=1}^{3+n_{s}} U_{si} \nu_{iL}\,, 
\label{005}
\end{equation}

where $n_{s}$ is the number of {\em sterile fields $\nu_{s}$}, the fields 
which 
do not enter into CC and NC interactions.
\end{enumerate}
In spite sterile neutrinos can not be detected via the standard 
weak interaction, there are different ways to search for the transition of flavor neutrinos into sterile states.

\begin{itemize}
\item

If it will be found
by the observation of neutrinos via the detection of NC processes that

\begin{equation}
\sum_{l'=e,\mu,\tau}P (\nu_{l}\to \nu_{l'})< 1   
\label{006}
\end{equation}
this would be an evidence of a transition of $\nu_{l}$
into sterile states.

\item

If we need more than 
two neutrino mass-squared differences in order to describe
experimental data, in this case we must assume that
more than three massive and mixed neutrinos exist
and there are transitions of flavor neutrinos into sterile states
(this will be the case, if the LSND result is confirmed).
\item
We can obtain an information about transition of flavor neutrinos into sterile states by the investigation of
matter effects.

\end{itemize}

{\em The smallness of neutrino masses can be naturally explained in the framework of a new physics.}

The most popular mechanism of the generation of small neutrino masses is
{\em the see-saw mechanism} \cite{see-saw}. This mechanism 
is based on the assumption that lepton number is violated 
by a right-handed Majorana mass term
at a scale M which is much larger than the electroweak scale.
For neutrino masses we have in this case

\begin{equation}
m_{i} \simeq \frac { (m_{f}^{i})^{2}} {\rm{M}} \ll m_{f}^{i}\,,
\label{007}
\end{equation}
where $m_{f}^{i}$ is mass of quark or lepton in $i$ family

If neutrino masses are of the see-saw origin in this case

\begin{itemize}
\item
Neutrinos with definite masses $\nu_{i}$ (i=1,2,3) are {\em Majorana particles}.
\item

Neutrino masses satisfy the hierarchy

$$ m_{1}\ll m_{2}\ll m_{3}\,.$$

\item

Heavy Majorana particles with masses $ \simeq\rm{M}$ must exist.
The existence of such particles could  provide a mechanism for an explanation
of the baryon asymmetry of the Universe \cite{buch}.

\end{itemize}

Smallness of neutrino masses can be due to the existence of {\em large extra dimensions} \cite{largedim}. 
In this case 
$\nu_{i}$ are {\em Dirac  particles}.

\item
\begin{center}
{\bf Neutrino oscillations}
\end{center}
\vskip0.5cm
In case of neutrino mixing, the state of flavor neutrino $\nu_{l}$
with momentum
$\vec p$ is a {\em coherent superposition} of the states of neutrinos with 
definite masses (see, \cite{BilP,BilGun})
\begin{equation}
|\nu_{l}> = \sum_{i} U_{li}^{*} |\nu_{i}> \,,
\label{008}
\end{equation}
where
$|\nu_{i}>$ is the state of neutrino with mass $m_{i}$, momentum $\vec p$,
energy $E_{i}= \sqrt{m_{i}^{2}+ p^{2}} \simeq p + \frac {m_{i}^{2}}{2p}$
and negative helicity. The equation (\ref{008})  is based 
on the smallness of neutrino mass-squared differences with respect to the square of neutrino energy.

The probabilities of the transition $\nu_\alpha \to \nu_{\alpha'}$
and $\bar\nu_\alpha \to\bar \nu_{\alpha'}$ are given by the following general expressions (see, \cite{BGG})
\begin{equation}
{\mathrm P}(\nu_\alpha \to \nu_{\alpha'}) =
|\delta_{{\alpha'}\alpha} +\sum_{i} U_{\alpha' i}  U_{\alpha i}^*
\,~ (e^{- i \Delta m^2_{i 1} \frac {L} {2E}} -1)|^2 \,.
\label{009}
\end{equation}
and 

\begin{equation}
{\mathrm P}(\bar\nu_\alpha \to \bar\nu_{\alpha'}) =
|\delta_{{\alpha'}\alpha} +\sum_{i} U_{\alpha' i}^*  U_{\alpha i}
\,~ (e^{- i \Delta m^2_{i 1} \frac {L} {2E}} -1)|^2 \,.
\label{010}
\end{equation}

In these expressions $L$ is the distance between a neutrino source and a neutrino detector and
$E$ is the neutrino energy.

Let us stress that 
\begin{itemize}
\item
The transition probabilities depend on $\frac {L} {E}$
\item
Neutrino oscillations  can be observed if 

$$
\Delta m^2_{i 1} \frac {L} {E}\gtrsim 1 $$

for at least one $ i$. In this condition $\Delta m^2_{i 1}$ is in $\rm{eV}^{2}$,
$L$ is in m (km) and $E$ is in MeV (GeV).
\end{itemize}

The typical values of
 $\frac {L} {E}$
are in the range
[$1-10^{3}$] in accelerator neutrino experiments, in the range
[$10^{2}-10^{5}$] in
reactor neutrino experiments,
in the range [$10-10^{4}$] in  atmospheric neutrino experiments
and in the range [$10^{10}-10^{11}$] in
solar neutrino experiments.

\item
\begin{center}
{\bf Oscillations between two types of neutrinos} 
\end{center}
\vskip0.5cm
 
 We will present here the standard formulas for the probabilities
of transitions between two types of neutrinos
($\nu_{\mu} \to \nu_{\tau}$ or  $\nu_{\mu} \to \nu_{e}$ etc).
From the general expressions Eq.(\ref{009}) 
and Eq.(\ref{010})
we have in this case

\begin{equation}
{\mathrm P}(\nu_\alpha \to \nu_{\alpha'}) ={\mathrm P}(\bar \nu_\alpha \to 
\bar \nu_{\alpha'})=
|\delta_{{\alpha'}\alpha} + U_{\alpha' 2}  U_{\alpha 2}^*
\,~ (e^{- i \Delta m^2 \frac {L} {2E}} -1)|^2 \,,
\label{011}
\end{equation}
where $\Delta m^2= m^2_{2}-m^2_{1} $. 

From this expression for
the appearance probability ($\alpha \not= \alpha'$)
we obtain 

\begin{equation}
{\mathrm P}(\nu_\alpha \to \nu_{\alpha'}) = {\mathrm P}(\bar \nu_\alpha \to 
\bar \nu_{\alpha'})
= \frac {1} {2} {\mathrm A}_{{\alpha'};\alpha}\,~
 (1 - \cos \Delta m^{2} \frac {L} {2E})\,.
\label{012}
\end{equation}
Here
$${\mathrm A}_{{\alpha'};\alpha}= 4\,~|U_{\alpha' 2}|^{2}
\,~|U_{\alpha 2}|^{2}\,.$$

The disappearance probability is given by

\begin{equation}
{\mathrm P}(\nu_\alpha \to \nu_\alpha) ={\mathrm P}(\bar \nu_{\alpha} \to
\bar \nu_{\alpha})=
 1 - \frac {1} {2}{\mathrm B}_{\alpha ; \alpha}\,~ 
(1 - \cos \Delta m^{2} \frac {L} {2E})\,,
\label{013}
\end{equation}
where
$${\mathrm B}_{{\alpha};\alpha}= 4\,~|U_{\alpha' 2}|^{2}\,~
(1-|U_{\alpha 2}|^{2})\,.$$

From the unitarity of the mixing matrix it follows that oscillation 
amplitudes in the disappearance and appearance channels are connected by the relation

\begin{equation}
{\mathrm B}_{\alpha ; \alpha}={\mathrm B}_{\alpha' ; \alpha'}=
{\mathrm A}_{{\alpha'};\alpha}=\sin^{2}  2\theta \,
\label{014}
\end{equation}
where $\theta$ is the mixing angle.

\end{enumerate}

\section{Evidence in favor of oscillations of atmospheric neutrinos}

Atmospheric neutrinos are produced mainly  in the decays of pions and
muons

$$\pi \to \mu + \nu_{\mu}\,~~ \mu  \to e + \nu_{\mu} + \nu_{e}\,. $$

In the Super-Kamiokande(Super-K) experiment \cite{AS-K} a large 
water Cherenkov detector (50 ktons of water) is used.
Electrons and muons, produced in the interaction of the atmospheric neutrinos with 
nuclei, are detected by 11200 photomultipliers via the observation of 
the Cherenkov light.

The observation of large zenith angle $\theta_{z}$ asymmetry of the high energy muon events
in the Super-K experiment constitute compelling evidence in favor of neutrino oscillations.

If there are no neutrino oscillations, the number of the detected electrons (muons) must satisfy the following symmetry relation

\begin{equation}
N_{l}(\cos\theta_{z})= N_{l}( -\cos \theta_{z})\,~~ (l=e,\mu)
\label{015}
\end{equation}
For the electron events 
a good agreement with this relation was observed. 

{\em A significant $\cos\theta_{z}$ asymmetry of the Multi-GeV
($E \geq 1.3\,~ \rm{GeV}$) muon events was observed in the Super-K experiment.}
For 
the ratio of total numbers of the up-going and the
down-going muons it was found

\begin{equation}
\left(\frac{U}{D}\right)_{\mu} = 0.54 \pm 0.04 \pm 0.01\,.
\label{016}
\end{equation}

Here U is the total number of up-going muons ($500\,~ \rm{km} \leq L 
\leq 13000 \,~ \rm{km}$) 
and D is the total number of down-going muons ($20\,~ \rm{km} \leq L \leq 500\,~ \rm{km}$).

The Super-K data and data of other atmospheric neutrino experiments
(SOUDAN 2 \cite{Soudan}, MACRO \cite{MACRO} ) are well described, 
if we assume that the two-neutrino $\nu_{\mu}\to
\nu_{\tau}$ oscillations take place.

 For the best-fit values of neutrino oscillation parameters
 from the combined fit of the Super-K data it was found 
\begin{equation}
\Delta m^{2}_{\rm{atm}} =  2.5 \cdot 10^{-3}\rm{eV}^{2}\,~~
 \sin^{2}2 \theta_{\rm{atm}} =1\,~ (\chi^{2}_{min}= 142.1 \,; 152 \rm{ d.o.f.})\,.
\label{017}
\end{equation}

\section {Evidence of oscillations of solar neutrinos} 

The event rates measured in all solar neutrino experiments 
(Homestake \cite{Cl},Kamiokande \cite{Kam},
 GALLEX-GNO \cite{GALLEX,GNO}, SAGE\cite{SAGE},
Super-K \cite{S-K} and SNO \cite{SNO,SNONC,SNOCC} ) are significantly smaller than the event rates predicted by the Standard Solar models
(SSM).
For the ratio R of the observed and predicted by SSM BP00 \cite{BPin}
event rates
it was found

\bea
 &&R = 0.34 \pm 0.03 \,~~~~ (\mathrm{Homestake})
\nonumber\\
&&R = 0.58 \pm 0.05 \,~~~~ (\mathrm{GALLEX-GNO})
\nonumber\\
 &&R= 0.60 \pm 0.05 \,~~~~ (\mathrm{SAGE})
\nonumber\\
 &&R = 0.459 \pm 0.017  \,~~ (\mathrm{Super-K})
\nonumber\\
&&R = 0.35\pm 0.03  \,~~~~  (\mathrm{SNO})
\nonumber
\eea
{\em The strong model independent evidence in favor of the transition of
solar $\nu_{e}$ into $\nu_{\mu}$ and $\nu_{\tau}$ was obtained  
after the results of the SNO experiment \cite{SNO,SNONC,SNOCC}}
were published.

The detector in the SNO experiment  is a 
heavy water Cherenkov detector 
(1 kton of $\rm{D}_{2}$O).
Solar neutrinos are observed via the detection of the 
CC reaction
\be
\nu_e + d \to e^{-}+ p +p\,,
\label{018}
\ee
the  NC reaction 

\be
\nu + d \to \nu+ n +p\,.
\label{019}
\ee
and the elastic scattering of neutrinos on electrons (ES)
\be
\nu + e \to \nu + e \,,
\label{020}
\ee

During the first 241 days of running in the SNO experiment $975.4 \pm 39.7$
CC events and $106.1 \pm 15.2$ ES events were observed.
The threshold for the detection of the electrons was 6.75 MeV.

The total CC event rate is given by

\be
R^{CC}=  <\sigma^{CC}_{\nu_{e}d}>\Phi_{\nu_{e}}^{CC}\,,
\label{021}
\ee
where $<\sigma^{CC}_{\nu_{e}d}>$ is cross section of the process
(\ref{018}),
averaged over initial spectrum of
$^{8}\rm{B}$ neutrinos, and $\Phi_{\nu_{e}}^{CC}$
is the flux of $\nu_e$ on the earth. We have
\be
\Phi_{\nu_{e}}^{CC} = <P(\nu_e \to\nu_e)>_{CC}\,~\Phi_{\nu_{e}}^{0}\,, 
\label{022}
\ee
where $\Phi_{\nu_{e}}^{0}$ is the total initial flux of  $\nu_e$
and $<P(\nu_e \to\nu_e)>_{CC}$ is the averaged $\nu_e$ survival probability.

From the CC event rate, measured in the SNO
 experiment, for the flux of $\nu_e$ on the earth the value

\be
(\Phi_{\nu_{e}}^{CC})_{SNO} =
({1.75^{+0.07}_{-0.07}\mbox{(stat.)}^{+0.12}_{-0.11}~\mbox{(syst.)}}) 
 \cdot 10^{6}\,~
cm^{-2}s^{-1} \,.
\label{023}
\ee
was found \cite{SNO}.

The first evidence in favor of the presence of 
$\nu_{\mu}$ and $\nu_{\tau}$ in the flux of the solar neutrinos
on the earth was obtained from the combination of the SNO and the Super-K
results.

In  the Super-K experiment \cite{S-K} solar neutrinos are observed via
the detection of the ES reaction (\ref{020}).
During 1258 days of running 
$18464 \begin{array}{c} +677 \\-590\end{array}$
events with the recoil electron energies in the range 5-20 $\rm{MeV}$ were observed.

Because of the high energy threshold
mostly 
neutrinos from the decay $^{8} B \to^{8} Be^{*} +  e^{+}+\nu_e $  
are detected in the Super-K and in the SNO experiments.\footnote{ The $hep$ neutrinos
give a small contribution to the event rates. For the flux of the 
$hep$ neutrinos the SSM BP00 value are usually used}  
Let us stress that the initial spectrum of  $^{8} B$ neutrinos  
is determined by 
the weak interaction and is known \cite{Ortiz}.

The total ES event rate is given by

\be
R^{ES}=  <\sigma_{\nu_{e}e}>\Phi_{\nu_{e}}^{ES} 
+ <\sigma_{\nu_{\mu}e}>\,\sum_{l= \mu,\tau}\Phi_{\nu_{l}}^{ES}
\,,
\label{024}
\ee
where $<\sigma_{\nu_{l}e}>$ is the cross section of the process 
$\nu_{l}e \to\nu_{l}e$ averaged over the initial spectrum of the 
$^{8}\rm{B}$ neutrinos and $\Phi_{\nu_{l}}^{ES}$ is the flux of 
$\nu_{l}$ on the earth measured via the observation of the $\nu_{l}-e$ scattering.

We can write the expression (\ref{024}) in the form
\be
R^{ES}=  <\sigma_{\nu_{e}e}>\Phi_{\nu}^{ES}\,.
\label{025}
\ee

Here

\be
\Phi_{\nu}^{ES}= \Phi_{\nu_{e}}^{ES} + \frac{<\sigma_{\nu_{\mu}e}>}
{<\sigma_{\nu_{e}e}>}\,~\sum _{l=\mu,\tau }\Phi_{\nu_{l}}^{ES}\,,
\label{026}
\ee
where 
$$\frac{<\sigma_{\nu_{\mu}e}>}{<\sigma_{\nu_{e}e}>}=0.154 \,.$$

The flux of $\nu_{\mu}$ and $\nu_{\tau}$ on the earth
enters into expression for the flux $\Phi_{\nu}^{ES}$
with the small coefficient 0.154.
This is connected with the fact the the cross section of 
the NC $\nu_{\mu}-e $ scattering is about 6 times smaller that the 
cross section of the NC+CC $\nu_{e}-e $ scattering.
Thus, the sensitivity to  $\nu_{\mu}$ and $\nu_{\tau}$ of experiments
in which solar neutrinos are observed through the detection of ES process
(\ref{020}) is significantly smaller than the sensitivity to $\nu_{e}$.

For the flux $\Phi_{\nu}^{ES}$ in the Super-K experiment 
it was found

\be
(\Phi_{\nu}^{ES})_{SK}= (2.32 \pm 0.03 \pm 0.08) 
\cdot 10^{6}cm^{-2}s^{-1}\,. 
\label{027}
\ee

No distortion of the spectra of electrons were observed in the Super-K and 
in the SNO experiments. From these data it follows that at high energies

$$P(\nu_e \to\nu_e)\simeq const\,.$$
and 
$$
\Phi_{\nu_{e}}^{CC}\simeq \Phi_{\nu_{e}}^{NC}\,.$$

Taking into account this relation from (\ref{023})
and (\ref{027}) for the flux of 
 $\nu_{\mu}$ and $\nu_{\tau}$ on the earth the value 

\be
\sum _{l=\mu,\tau }\Phi_{\nu_{l}}^{ES} = 3.69 \pm 1.13\cdot 10^{6}cm^{-2}s^{-1}
\label{028}
\ee
was found \cite{SNO}.

Thus from the comparison of the neutrino fluxes measured in the Super-K
and SNO experiment the first model independent evidence (at 3 $\sigma$ level)
of the presence of  $\nu_{\mu}$ and $\nu_{\tau}$ in the flux of the solar
neutrinos on the earth was obtained.

Recently the first results of the observation of the solar neutrinos through the detection of the NC reaction (\ref{019}) were published by the SNO collaboration \cite{SNONC}. New CC and ES data were also obtained \cite{SNOCC}.
During 306.4 days of running $576.5^{+49.5}_{-48.9}$ NC events,
 $1967^{+61.9}_{-60.9}$  
CC events and  $263.6^{+26.4}_{-25.6}$ 
ES events were recorded. 
The kinetic energy threshold for the detection of electrons was equal to
5 MeV. The NC threshold is 2.2 MeV.

All flavor neutrinos are recorded by 
the observation of the solar neutrinos through the detection of the NC process (\ref{019}).
Assuming $\nu_{e}-\nu_{\mu}- \nu_{\tau}$
universality of the NC for the total NC event rate we have

\be
R^{NC}=  <\sigma^{NC}_{\nu d}>\Phi_{\nu}^{NC}\,.
\label{029}
\ee

Here  $<\sigma^{NC}_{\nu d}>$ is the cross section of the process
(\ref{019}), averaged over the known initial spectrum of the 
$^{8}\rm{B}$ neutrinos, and 
\be
\Phi_{\nu}^{NC} = \Phi_{\nu_{e}}^{NC} + \sum_{l=\mu,\tau}\Phi_{\nu_{l}}^{NC}
\label{030}
\ee
is the flux of all flavor neutrinos on the earth. In the SNO experiment 
\cite{SNONC,SNOCC} for the fluxes  $(\Phi_{\nu}^{NC})$,
 $(\Phi_{\nu_{e}}^{CC})$ 
and 
$(\Phi_{\nu}^{ES})$ the following values were found

\be
(\Phi_{\nu}^{NC})_{SNO} =({5.09^{+0.44}_{-0.43}\mbox{(stat.)}^{+0.46}_{-0.43}~\mbox{(syst.)}} ) \cdot 10^{6}\,~
cm^{-2}s^{-1} \,,
\label{031}
\ee

\be
(\Phi_{\nu}^{CC})_{SNO} =({1.76^{+0.06}_{-0.05}\mbox{(stat.)}^{+0.09}_{-0.09}~\mbox{(syst.)}} ) \cdot 10^{6}\,~
cm^{-2}s^{-1} \,,
\label{032}
\ee

$$
(\Phi_{\nu}^{ES})_{SNO} =({2.39^{+0.24}_{-0.23}\mbox{(stat.)}^{+0.12}_{-0.12}~\mbox{(syst.)}} ) \cdot 10^{6}\,~
cm^{-2}s^{-1} \,.$$

Taking into account that 

$$\Phi_{\nu_{e}}^{NC}\simeq \Phi_{\nu_{e}}^{CC}$$

from (\ref{030}), (\ref{031}) and (\ref{032}) 
for the
flux of 
$\nu_{\mu}$ and $\nu_{\tau}$
on the earth the following value 
was found \cite{SNONC}

\be
\sum_{l=\mu,\tau}\Phi_{\nu_{l}}^{NC} =(3.41^{+0.45}_{-0.45}\mbox{(stat.)}^{+0.48}_{-0.45}~\mbox{(syst.)})\cdot 10^{6}\,~
cm^{-2}s^{-1} \,.
\label{033}
\ee

Thus, the observation of the solar neutrinos  
through the  simultaneous detection of the
CC process (\ref{018}) and the NC process (\ref{019}) 
allowed to obtain the {\em direct model independent evidence 
(at $5.3\,~ \sigma$ level)
of the presence of
$\nu_{\mu}$ and $\nu_{\tau}$ in the flux of the solar neutrinos on the earth}.

Let us notice that the total flux of all flavor neutrinos on the earth,
measured in the SNO experiment, is in agreement with the flux of the $^{8} B$
neutrinos predicted by the SSM BP00 \cite{BPin}

$$(\Phi_{\nu_{e}})_{SSM}=5.05 (1 \pm 0.18)\cdot 10^{6}\,~
cm^{-2}s^{-1} \,.$$

The data of all solar neutrino experiments can be described 
under the assumption of the two- neutrino
oscillations which are characterized
by two oscillation parameters  $\Delta m^{2}_{\rm{sol}}$ and $\tan^{2}\theta_{\rm{sol}}$. It was assumed also that solar neutrino fluxes are given by
the SSM.

Before the Super-K 
measurement of the spectrum of the recoil electrons, 
from the fit of the event rates measured in the Homestake, GALLEX, SAGE and 
Super-K experiments, several allowed regions (solutions) in the plane of the oscillation parameters were found:
large mixing angle  MSW solutions LMA and LOW, 
small mixing angle MSW solution SMA and vacuum solution VO. The situation changed after 
the day and night recoil electron spectra were measured in the Super-K experiment. The analysis of these and other solar neutrino data allowed to conclude that the most plausible allowed regions are LMA and LOW \cite{Fukuda}.
Analysis of the solar neutrino data, that were done after the first SNO data appeared \cite{Bahcall,Fogli}, confirm this conclusion.

The analysis of the new SNO data \cite{SNOCC,Barger} strongly favor
the LMA solution.
In \cite{SNOCC} in addition to the oscillation parameters
 $ \Delta m^{2}_{\rm{sol}}$ and 
$\tan^{2}\theta_{\rm{sol}}$ also the total initial flux of the $^{8}\rm{B}$ neutrinos $\Phi_{\nu_{e}}^{0}$ was considered as a free parameter.

In the case of the LMA allowed region
for the 
best-fit values of the parameters
it was found     

$$
 \Delta m^{2}_{\rm{sol}} = 5 \cdot 10^{-5}\,\rm{eV}^{2};\,~
 \tan^{2}\theta_{\rm{sol}} = 3.4 \cdot 10^{-1}\,~
\Phi_{\nu_{e}}^{0}= 5.86\cdot 10^{6}cm^{-1}s^{-1}$$
$$(\chi^{2}_{\rm{min}}= 57.0;\,~ 72 \,~ d.o.f.)$$    

The LOW solution appears at 99.5\% CL. For the best-fit values of the parameters it was obtained    
$$
 \Delta m^{2}_{\rm{sol}} = 1.3 \cdot 10^{-7}\,\rm{eV}^{2};\,~
 \tan^{2}\theta_{\rm{sol}} = 0.5 \cdot 10^{-1}\,~
\Phi_{\nu_{e}}^{0}= 4.95\cdot 10^{6}cm^{-1}s^{-1}$$
$$(\chi^{2}_{\rm{min}}= 67.7; \,~  72 \,~d.o.f.)$$

\section {Implications of the results of the SNO and Homestake experiments
 for BOREXINO}

We will consider here\cite{BLPF} implications  that can be
inferred  from the
results of the
SNO
and Homestake
experiments    
for the future BOREXINO experiment \cite{BOR}

Let us obtain  the contribution of the medium energy $^7 Be$, CNO and pep
neutrinos to the chlorine event rate 

\be
 R_{Cl} = (2.56 \pm 0.16 \pm 0.16) \,~~\rm{SNU}\,,
\label{034}
\ee
measured in the Homestake experiment \cite{Cl}.

The main contribution to the chlorine event rate
is due to
the $^8 B$ and  the $^7 Be$ neutrinos.
According to SSM BP00 \cite{BPin}, the contributions of $^8 B$ and  $^7 Be$ neutrinos to the chlorine event rate are equal to, respectively, 5.9 SNU and 1.15 SNU.
Less important but sizable contribution comes from
the CNO and the pep neutrinos (0.7 SNU).  

Using the value of the total flux of $^8 B$ neutrinos, measured 
in the SNO experiment,
we calculate first the contribution of the  $^8 B$
neutrinos to the chlorine event rate.
We have
\be
 (R_{Cl}^{^8B})_{SNO} = \int_{E_{th}}\sigma_{\nu_{e}Cl}(E)\,~
X^{^8B}(E)\,~\Phi_{\nu_{e}}^{SNO}\,dE\,, 
\label{035}
\ee
where $\sigma_{\nu_{e}Cl}(E)$ is the cross section of the process
$\nu_{e}+ ^{37}Cl\to e^{-}+^{37}Ar$ and $X(E)$ is the normalized initial spectrum of the $^8 B$ neutrinos. Using (\ref{023}) and (\ref{035}), for the  
 contribution of the  $^8 B$
neutrinos to the chlorine event rate we obtain

\be
(R_{Cl}^{^8B})_{SNO} = (2.00 \pm 0.17)\,\rm{SNU}.
\label{036}
\ee
Further, from (\ref{034}) and (\ref{036}) we find that 
the contribution of $^7 Be$, CNO and pep neutrinos to the chlorine event rate is equal to

\be
 R_{Cl}^{^7 Be,CNO,pep} = (0.56 \pm 0.29) \,~\rm{SNU}\,.
\label{037}
\ee
This value is more than 4$\sigma$ smaller than SSM BP00 prediction (1.8 SNU).

In order to determine from (\ref{037}) the flux of the $^7 Be$ neutrinos on the earth 
we must take into account the relatively small contribution 
to the chlorine event rate
 of the CNO and the pep neutrinos. 
In the calculation of this contribution 
LMA (and LOW) solutions were used in \cite{BLPF}.
For the flux of 
the ${^7 Be}$ neutrinos on the earth it was found

\be
 \Phi_{\nu_{e}}^{^7Be} = (1.19 \pm 1.12)\cdot 10^{9}\,\rm{cm}^{-2}\rm{s}^{-1}\,~~~ LMA 
\label{038}
\ee
and
\be
 \Phi_{\nu_{e}}^{^7Be} = (1.00 \pm 1.08)\cdot 10^{9}\,\rm{cm}^{-2}\rm{s}^{-1}\,~~~ LOW\,. 
\label{036}
\ee
Thus, the flux of the ${^7 Be}$ neutrinos on the earth, that can be obtained from the results of the SNO and the Homestake experiments, is  
significantly smaller than the flux predicted by the SSS BP00
$$
( \Phi_{\nu_{e}}^{^7Be})_{SSM} = (4.77 \pm 0.47)\cdot 10^{9}\,\rm{cm}^{-2}\rm{s}^{-1}\,. $$

Finally, for the BOREXINO event rate from (\ref{035}) and (\ref{036})
it can be obtained, correspondingly

$$ R_{\rm{Borexino}} =24.4 \pm 8.9\,~ \rm{events/day}\,~
 R_{\rm{Borexino}} =22.8 \pm 8.6\,~ \rm{events/day}$$

These values are compatible with LMA and LOW prediction (30.7 events/day
and 29.0 events/day). The SSM BP00 for the BOREXINO event rate predicts
55.2 events/day.

\section{Neutrino oscillations in the  atmospheric 
range  of $\Delta m^{2}$
in the framework
of three-neutrino mixing}

In the framework of the three-neutrino mixing we will consider here neutrino oscillations 
in the atmospheric and  
the accelerator (reactor)
long baseline (LBL) neutrino experiments.
Due to the hierarchy of neutrino mass-squared differences 
\be
\Delta m^{2}_{21}\ll\Delta m^{2}_{31}
\label{040}
\ee
we can neglect the contribution of the $i=2$ term in the expression
(\ref{009}) for the transition probability.
For probability of the transition $\nu_\alpha \to \nu_{\alpha'}$
we have in this case

\be
{\mathrm P}(\nu_\alpha \to \nu_{\alpha'}) \simeq
|\delta_{{\alpha'}\alpha} + U_{\alpha' 3}  U_{\alpha 3}^*
\,~ (e^{- i \Delta m^2_{31} \frac {L} {2E}} -1)|^2 
\label{041}
\ee
From the comparison of  (\ref{011}) and (\ref{041})
it is obvious that transition probabilities in the atmospheric and LBL experiments have
in the case of the hierarchy (\ref{040}) the
two-neutrino form (see, \cite{BGG})

\begin{equation}
{\mathrm P}(\nu_\alpha \to \nu_{\alpha'}) =
 \frac {1} {2} {\mathrm A}_{{\alpha'};\alpha}\,~
 (1 - \cos \Delta m^{2}_{31} \frac {L} {2E})\,~~~(\alpha \not= \alpha')
\label{042}
\end{equation}

and

\begin{equation}
{\mathrm P}(\nu_\alpha \to \nu_\alpha)=
 1 - \frac {1} {2}{\mathrm B}_{\alpha ; \alpha}\,~ 
(1 - \cos \Delta m^{2}_{31} \frac {L} {2E})\,.
\label{043}
\end{equation}
The oscillation amplitudes are given by the expressions

\be
{\mathrm A}_{{\alpha'};\alpha}= 4\,~|U_{\alpha' 3}|^{2}\,~|U_{\alpha 3}|^{2}
 \,~~~{\mathrm B}_{\alpha ; \alpha}=4\,~|U_{\alpha 3}|^{2}
\,~(1 -|U_{\alpha 3}|^{2}).
\label{044}
\ee

From the unitarity of the mixing matrix it follows that the oscillation amplitudes in the disappearance and appearance channels are connected by the relation

\be
{\mathrm B}_{\alpha ; \alpha}= \sum_{\alpha'\not=\alpha}{\mathrm A}_{{\alpha'};\alpha}\,.
\label{045}
\ee

Let us stress that the expressions (\ref{042}) and (\ref{043}) describe 
neutrino oscillations in all three oscillation channels:
$\nu_{\mu} \to \nu_{\tau}$, 
$\nu_{\mu} \to \nu_{e}$ and $\nu_{e} \to \nu_{\tau}$.

Because of the unitarity relation  $\sum_{\alpha} |U_{\alpha 3}|^{2}= 1$,
transition probabilities (\ref{042}) and (\ref{043})
depend on three parameters.
We can choose  
\be
\Delta m^2_{31}\,~~ \tan^2\theta_{23}\,~~ |U_{e 3}|^2\,.
\label{046}
\ee
For the oscillation amplitudes we have

\be
{\mathrm A}_{\tau;\mu}= ( 1 - |U_{e3}|^2)^2\,\sin^{2}2\theta _{23}\,~~
{\mathrm A}_{e;\mu}=4\,|U_{e3}|^2 \,( 1 - |U_{e3}|^2)\,\sin^{2}\theta _{23}\,.
\label{047}
\ee

The Super-K data are well described, if we assume that

 $$|U_{e3}|^2 =0 $$
In this case in the leading approximation
pure $\nu_{\mu} \to \nu_{\tau}$ oscillations will take place and

$${\mathrm A}_{\tau;\mu}={\mathrm B}_{\mu;\mu}= \sin^{2}2\theta _{23}
\equiv \sin^{2}2\theta _{\rm{atm}}$$

The Super-K best-fit values of the parameters $\Delta m^{2}_{\rm{atm}}$
and $\sin^{2}2\theta _{\rm{atm}}$ are given by (\ref{017}).

From a more general analysis of the Super-K data with three parameters
(\ref{046}) the following upper bound on the parameter $|U_{e3}|^2$ was found 
\cite{Kaji}
      $$|U_{e3}|^2 \lesssim 0.35 $$

The most 
stringent bound on $|U_{e3}|^2 $ can be obtained 
from the results of the  LBL reactor CHOOZ \cite{CHOOZ2} and Palo Verde 
\cite{PaloV} experiments
in which the disappearance of the reactor $\bar \nu_e $'s in the atmospheric range of neutrino mass-squared difference was  searched for.
No indications in favor of  $\bar \nu_e $ disappearance were found in these experiments. 

From the exclusion plot, obtained from the analysis of the data of
 these experiments, we have the bound

\be
{\mathrm B}_{e;e}=4\, |U_{e3}|^2 \,( 1 - |U_{e3}|^2)
\leq{\mathrm B}^{0} _{e ; e}\,,
\label{048}
\ee
where ${\mathrm B}^{0} _{e ; e}$
depends on $\Delta m^{2}_{31}$.

 For the parameter $|U_{e 3}|^{2}$
 from (\ref{048}) we have

\be
|U_{e 3}|^{2} \leq  
\frac{1}{2}\,\left(1 - \sqrt{1-{\mathrm B}_{e;e}^{0} }\right)\simeq 
\frac{{\mathrm B}_{e;e}^{0}}{4}
\label{049}
\ee

or
\be
|U_{e 3}|^{2} \geq
\frac{1}{2}\,\left(1 + \sqrt{1-{\mathrm B}_{e;e}^{0} }\right)\simeq
1-\frac{{\mathrm B}_{e;e}^{0}}{4}
\label{050}
\ee
The second possibility is excluded by the solar neutrino data. In fact,
as we will see later, if the parameter $|U_{e 3}|^{2}$ is close to one then
the probability of solar $\nu_{e}$ to survive in the whole
energy region is is also close to one, in obvious contradiction with 
the solar neutrino data.
Thus, the parameter $|U_{e 3}|^{2}$ is small and satisfies the inequality
(\ref{049}).

From  the exclusion curve, obtained from the results of the CHOOZ experiment,
at  
$\Delta m^2_{31}= 2.5\,~10^{-3}\rm{eV}$
(the Super -K best-fit value) we have

\be
|U_{e 3}|^{2}\leq 3.7 \,~10^{-2}
\label{051}
\ee

The value of $|U_{e 3}|^{2}$ is very important for 
the future high precision neutrino experiments. In particular, the possibilities to
investigate the effects of CP violation in the lepton sector
at the Super Beam Facility (\cite{jhf}) and at the future Neutrino Factory (see, \cite{Cline})
depend on the value of this parameter.

\section{A bound on $|U_{e 3}|^{2}$ from three- neutrino analysis of the CHOOZ data}
The bound (\ref{051}) was obtained for the small values of $\Delta m^2_{21}$.
In this case
the contribution of the $i=2$ term 
in Eq.(\ref{010}) for the transition probability  can be neglected. As we stressed before, the 
most plausible solution of the solar neutrino problem is the LMA solution.
From the global analyses of the solar neutrino data it follows
that neutrino oscillation parameters in the case of the LMA solution can 
be varied in rather wide 
ranges. For example, in \cite{Bahcall} it was found

\be
2\cdot 10^{-5} \lesssim \Delta{m}^2_{\rm{sol}}\lesssim
 6\cdot 10^{-4}\rm{eV}^{2}\,;~~
   0.4 \lesssim  \tan^{2}\theta _{\rm{sol}}\lesssim 1 
\label{052}
\ee
If $\Delta m_{\rm{sol}}^{2}\geq 10^{-4}\rm{eV}^{2}$,
in this case the phase $\Delta m_{\rm{sol}}^{2}\frac{L}{2E}$ is not small and
the contribution of $i=2$ term in the Eq. (\ref{010})  
 can be important.

From (\ref{009}) for the $\bar \nu_{e}$ survival probability 
we can obtain the following exact expression \cite{BNP}

\begin{eqnarray}
\lefteqn{P({\bar \nu_e}\to{\bar \nu_e})} \nonumber\\
&& =\,~~ 1 - 2 \, |U_{e 3}|^2 \left( 1 - |U_{e 3}|^2 \right)
\left( 1 - \cos \frac{ \Delta{m}^2_{31} \, L }{ 2 \, E } \right)
\nonumber \\
&& -\,~~\frac{1}{2} \,(1- |U_{e 3}|^2)^{2}\sin^{2}2\,\theta _{\rm{12}}
\,
\left( 1 - \cos \frac{ \Delta{m}^2_{\rm{21}} \, L }{ 2 \, E } \right)\nonumber   \\
& & +\,~~ 2\, |U_{e 3}|^2 \,(1- |U_{e 3}|^2)\,\sin^{2}\theta _{\rm{12}}\,
\times 
\nonumber\\ 
& &\times \left(
\cos
\left( \frac
{\Delta{m}^2_{31} \, L }{ 2 \, E} - \frac {\Delta{m}^2_{\rm{21}} \, L }{ 2 \,
E}\right)
-\cos \frac {\Delta{m}^2_{31} \, L }{ 2 \, E} \right) 
\label{053}
\end{eqnarray}

In \cite{BNP} the results of the three-neutrino analysis of the CHOOZ
data were presented. It was assumed that $\Delta{m}^2_{\rm{21}}=\Delta{m}^2_{\rm{sol}}$
and $\tan^{2}\theta _{\rm{12}}=\tan^{2}\theta _{\rm{sol}}$ and that
the values of the parameters $\Delta{m}^2_{\rm{sol}}$ and 
$\tan^{2}\theta _{\rm{sol}}$ are in the LMA allowed region.
It was shown  that
at  $\Delta{m}^2_{\rm{sol}}\leq 2\cdot 10^{-4}\rm{eV}^{2}$
the three-neutrino CHOOZ bound on $|U_{e 3}|^{2}$
is practically the same as for the two-neutrino one. 
At larger values of $\Delta{m}^2_{\rm{sol}}$ a more stringent bound was obtained. For example,
at $\Delta{m}^2_{31}= 2.5 \cdot 10^{-3}\rm{eV}^{2}$ and 
$\sin^{2}\theta _{\rm{sol}}= 0.5 $ it was found
$$
|U_{e 3}|^{2}\leq 3 \,~10^{-2}\,;~~~
|U_{e 3}|^{2}\leq 2 \,~10^{-2}\,,$$
correspondingly, at
$$
\Delta{m}^2_{\rm{sol}}= 4\cdot 10^{-4}\rm{eV}^{2}\,;~~~
\Delta{m}^2_{\rm{sol}}= 6\cdot 10^{-4}\rm{eV}^{2}\,.$$

\section { Neutrino oscillations in the solar range of $\Delta{m}^2$ 
in the framework of the three-neutrino mixing}

The 
$\nu_{e}$ survival probability in vacuum can be written in the form

\be
{\mathrm P}(\nu_e\to\nu_e)
=
\left|
\sum_{i=1, 2}| U_{e i}|^2 \, 
 e^{ - i
 \, 
\Delta{m}^2_{i1} \frac {L}{2 E} }
 + | U_{e 3}|^2  \, 
 e^{ - i
 \, 
\Delta{m}^2_{31} \frac {L}{2 E} }\,\right|^2
\label{054}
\ee

Because of the 
hierarchy of the neutrino mass squared differences,
the interference between the first and the second terms in Eq.(\ref{051}) 
disappears after the averaging over neutrino energy, distance etc.
For the averaged survival probability we have

\be
{\mathrm P}(\nu_{e}\to\nu_{e})=|U_{{e} 3}|^{4}+ (1-|U_{{e} 3}|^{2})^{2}\,~
P^{(1,2)}(\nu_{e}\to\nu_{e})\,,
\label{055}
\ee

where 

\begin{equation}
{\mathrm P}^{(1,2)}(\nu_e \to \nu_e) =
 1 - \frac {1} {2}\sin^{2}2\,~\theta_{12}\,~ 
(1 - \cos \Delta m^{2}_{12} \frac {L} {2E})
\label{056}
\end{equation}
is the two-neutrino $\nu_{e}$ survival probability in vacuum.

The expression (\ref{055}) is also valid in the case of matter \cite{Schramm}.
In this case $P^{(1,2)}(\nu_{e}\to\nu_{e})$ is
the $\nu_{e}$ survival probability in matter
that can be obtained from the two-neutrino evolution equation, in which 
the usual interaction Hamiltonian is 
multiplied by $(1-|U_{{e} 3}|^{2})$.

Taking into account the bound (\ref{048}), we can neglect $|U_{{e} 3}|^{2}$
in the expression (\ref{052}). We have
\be
{\mathrm P}(\nu_{e}\to\nu_{e})\simeq P^{(1,2)}(\nu_{e}\to\nu_{e})
\label{057}
\ee

Thus, because of the neutrino mass-squared hierarchy and smallness of the parameter  $|U_{{e} 3}|^{2}$, oscillations in the solar range of $\Delta{m}^2$ are 
decoupled from oscillations in the atmospheric range of $\Delta{m}^2$
(see,\cite{BGun}).  
The $\nu_{e}$ survival probability
depends in this case only on parameters
$\Delta m^{2}_{21}$ and $ \tan^{2}\theta _{12}\,.$

If oscillation parameters $\Delta m^{2}_{12}$ and $\sin^{2}2\,~\theta_{12}$
are  in the  LMA allowed region {\em neutrino oscillations in
the solar range of neutrino mass squared difference
can be observed in reactor experiments with a distance between
a reactor and a detector about 100 km.}
The first such experiment KamLAND \cite{KamLAND} started in January 2002 in the Kamiokande mine in Japan.
The reactor
$\bar \nu_{e}$'s from several reactors at
 the average distance about 170 km from the Kamiokande mine are detected
in the  KamLAND experiment. 
About 700 events/kt/year is expected in the case of no oscillations.
After three years of running the whole LMA region of the neutrino oscillation parameters will be investigated in the  KamLAND experiment.

If it will be confirmed that neutrino oscillation parameters
$\Delta m^{2}_{12}$ and $\sin^{2}2\,~\theta_{12}$
can be  determined in reactor experiments,
{\em a new way of the investigation of the sun will be open}. Solar fluxes from different reactions 
can be directly measured in solar neutrino experiments.
Such measurements could provide a model independent check of the SSM.

\section { CONCLUSION}

There exist at present
compelling evidences of neutrino oscillations,
obtained in experiments with neutrinos from natural sources: in the 
atmospheric and in the solar neutrino experiments.

Oscillations in {\em the atmospheric range of $\Delta{m}^2$ }
are studied at present in the accelerator 
long baseline experiment K2K \cite{K2K}. In future
experiments
 MINOS \cite{ MINOS}, OPERA \cite{ OPERA}, ICARUS \cite{ICARUS} and others 
neutrino oscillations in the the atmospheric range of $\Delta{m}^2$
will be investigated in detail. 
In a more remote future in 
neutrino experiments at the Super Beam facility \cite{jhf} and at the Neutrino Factory 
(see, \cite {Cline})
neutrino oscillation parameters $|U_{e 3}|^{2}$,
$\Delta m^{2}_{31}$ and $\tan^{2}2\,~\theta_{23}$
will be measured with a high precision. In these experiments
CP violation in the lepton sector and other important features of
neutrino mixing will be investigated.

Oscillations in {\em the  solar range of $\Delta{m}^2$ }
will continue to be studied in the solar neutrino experiments. 
New solar experiments BOREXINO \cite{BOR} will be started soon.
If the KamLAND experiment will confirm that neutrino oscillation parameters
are in the LMA allowed region, a new way of 
the investigation of the sun will be open.

The MiniBooNe experiment in about 2 years will confirm or
refute the LSND claim. If the  LSND result will be confirmed,
existing neutrino oscillation data can not be described in the framework of 
the minimal scheme with three massive and mixed neutrinos.
We have to assume in this case that at least four mixed neutrinos with small masses exist
in nature.

There are, however, the fundamental problems of neutrino mixing 
that can not be solved by the neutrino oscillation experiments.

\begin{enumerate}

\item

{\em What is the nature of the massive neutrinos. Are they Dirac or Majorana particles?}

The experiments on the investigation of neutrinoless double $\beta $- decay 
could answer this question. Neutrino masses and elements of the mixing matrix enter in the matrix element of the  neutrinoless double $\beta $- decay in 
the form of the 
effective Majorana mass

 $$<m> =\sum_{i} \,U_{ei}^{2}\,m_{i}$$

The upper bound of $|<m>|$ can be found from the neutrino
oscillation data, if we 
make some assumptions on the neutrino mass spectrum (see, \cite{BGGKP}).
For example,
for the case of neutrino mass hierarchy $|<m>| \leq 5\cdot10^{-3} \rm{eV}$.

From the existing experimental data the
following upper bound
$$|<m>| \leq (0.2-0.6) \,~\rm{eV}$$
was obtained \cite{HM,IGEX}.
In the future experiments GENIUS , CUORE , MAJORANA , EXO \cite{futdouble}
and others the sensitivity 
$$|<m>|\simeq 10^{-2}\rm{eV}$$
will be reached.
\item
{\em What is the value of the minimal neutrino mass $m_{1}$?}

The best bound on the value of the minimal neutrino mass $m_{1}$
was obtained from the experiments on the measurement of the high energy part of the $\beta $- spectrum of $^{3}H$

$$m_{1}\leq 2.2\,\rm{eV\,~~} \rm{Mainz}\,~ \cite{Mainz}\,~~ 
m_{1}\leq 2.5\,\rm{eV}\,~~\rm{Troitsk}\,~ \cite{Troitsk}$$.

In the future experiment KATRIN \cite{Katrin} the sensitivity 

$$m_{1}\simeq (0.3-0.4)\,~\rm{eV}$$
is planned to be achieved.

\end{enumerate}

It is a pleasure for me to acknowledge support of the ``Programa de Profesores Visitantes de IBERDROLA de Ciencia y Tecnologia''.

\end{document}